\documentclass[journal=jpcbfk,manuscript=article]{achemso}
\usepackage{longtable}
\usepackage{multirow}
\usepackage{amsmath}
\usepackage[usenames,dvipsnames]{color}
\usepackage{soul}
\usepackage{threeparttable}
\usepackage{xr}
\usepackage{graphicx}
\usepackage{amsmath,amssymb}
\usepackage{caption}
\usepackage{color}
\usepackage{dcolumn}
\usepackage{bm}
\usepackage{float}
\usepackage{subcaption}
\usepackage{textcomp}
\usepackage{url}
\usepackage[normalem]{ulem}

\usepackage[unicode=true,
            bookmarks=true,
            bookmarksnumbered=true,
            bookmarksopen=true,
            bookmarksopenlevel=2,
            breaklinks=false,
            pdfborder={0 0 1},
            backref=false,
            colorlinks=true,
            hidelinks
            ]{hyperref}

\makeatletter

\author{Kai T\"opfer} \affiliation[University of
  Basel]{Department of Chemistry, University of Basel,
  Klingelbergstrasse 80 , CH-4056 Basel, Switzerland.}
\author{Meenu Upadhyay} \affiliation[University of
  Basel]{Department of Chemistry, University of Basel,
  Klingelbergstrasse 80 , CH-4056 Basel, Switzerland.}
\author{Markus Meuwly} \affiliation[University of Basel]{Department of
  Chemistry, University of Basel, Klingelbergstrasse 80 , CH-4056
  Basel, Switzerland.}  \email{m.meuwly@unibas.ch}

\title{Quantitative Molecular Simulations}

\begin{document}

\begin{abstract}
All-atom simulations can provide molecular-level insights into the
dynamics of gas-phase, condensed-phase and surface processes. One
important requirement is a sufficiently realistic and detailed
description of the underlying intermolecular interactions. The present
perspective provides an overview of the present status of quantitative
atomistic simulations from colleagues' and our own efforts for gas-
and solution-phase processes and for the dynamics on
surfaces. Particular attention is paid to direct comparison with
experiment. An outlook discusses present challenges and future
extensions to bring such dynamics simulations even closer to reality.
\end{abstract}

\section{Introduction}
In principle, quantum mechanics (QM) can exactly describe the
energetics of chemical systems of any size. However, despite the ever
increasing compute power available, determining the interactions and
forces for large systems required to follow their molecular dynamics
becomes computationally prohibitive. This is due to the unfavourable
scaling of electronic structure calculations with the number of
electrons and the large number of basis functions required for
accurately solving the Schr\"odinger equation although linear scaling
methods provide some remedies for such
disadvantages.\cite{bowler:2012} A quantum mechanical treatment for
the nuclear degrees of freedom is even more computationally demanding
and typical system sizes contain several 10 atoms for which rigorous
calculations are feasible.\cite{meyer:2009}\\

\noindent
On the other hand, empirical molecular mechanics (MM) energy functions
(``force fields'') are computationally advantageous to evaluate and
propagating the dynamics based on Newton's equations of motion allows
to access long time scales for large molecular systems. In this
context, ``long time scale'' is sub-microsecond and ``large systems''
means $10^9$ atoms for which performance of up to 8 ns/day can be
achieved.\cite{feig:2021} Such molecular dynamics (MD) simulations
have been used to investigate processes ranging from protein
folding,\cite{shaw:2014} ligand binding\cite{simonson:2002}, crowding
in cellular environments\cite{feig:2019} to characterizing
spectroscopic properties of solutes and peptides and reactions in the
gas phase and in solution.\cite{MM.jcp:2020} However one of the
challenges remains to develop suitable energy functions that retain
the precision of the QM methods they are often based on and that are
suitable to follow bond breaking and bond formation.\\

\noindent
One- or multi-dimensional vibrational spectroscopy is a powerful means
to characterize the structural dynamics of complex
systems.\cite{pines:2005,2dir:2011} Experiments greatly benefit and
often require accompanying MD simulations for their molecular-level
interpretation. Such atomistic simulations rely on the bonded and
non-bonded interactions to be described in a meaningful way for making
direct contact with experiments. Traditionally, empirical energy
functions use harmonic springs for chemical bonds and valence angles,
periodic functions for dihedrals, an atom-centered point charge-based
model for charges and a Lennard-Jones representation for van der Waals
interactions together with additional, more purpose-tailored
terms.\cite{mackerell2004} For applications in spectroscopy the
chemical bonds (stretching vibration) need to be described somewhat
more realistically to account for mechanical anharmonicity for which
Morse-oscillators are often sufficient because experiments at ambient
temperatures are not sensitive to highly vibrationally excited
states. Nevertheless, there is scope to use more accurate
representations, for example based on machine learning-type
approaches, in particular if reference data from high-level electronic
structure calculations are available. For the non-bonded interactions
- which include electrostatic and van der Waals terms - more
physics-based models that go beyond the standard representations have
been developed.\\

\noindent
The first-order treatment of the electrostatic interaction is based on
atom-centered point charges for Coulomb interactions. Such pair
interactions can be rapidly computed but lack the accuracy for
describing anisotropic contributions to the charge
density.\cite{Stone2013} Including higher-order atomic multipoles
improves the accuracy but at the expense of increased computational
cost and implementation
complexity.\cite{Handley2009,MM.mtp:2013,Devereux2014,bereau:2016}
Accounting for polarizability is another contribution that has been
recently included in empirical force fields and shows much promise for
further improvements of the computational models.\cite{ren:2019} From
an empirical force field perspective the van der Waals interactions
are often represented as Lennard-Jones terms with {\it ad hoc}
(Lorentz-Berthelot) combination rules. Alternative and potentially
improved representations are the buffered 14-7
parametrization\cite{halgren:1992} and modified combination
rules\cite{mason:1988,millie:2001}\\

\noindent
Current methods to investigate reactive systems in the gas- and
condensed-phase include mixed quantum mechanical/molecular mechanics
(QM/MM)\cite{Warshel76EVB,Alagona1986,Field1990}, reactive force
fields ReaxFF\cite{goddard01reaxff} and reactive MD
(RMD)\cite{nutt-biophys-06}. The general ansatz for reactive force
fields is to describe reactant and product states with a separate
energy function and to connect the two either by mixing functions or
by diagonalizing an $n \times n$ matrix, where $n$ is the number of
states considered. This reactive part is embedded in an environment
that is described by a more empirical energy function, akin to mixed
QM/MM calculations and simulations. In recent years, a variety of
sophisticated Machine Learning (ML) potentials were developed to
accurately represent \emph{ab initio} results from high level
reference calculations (see Ref. \citenum{Chmiela2018,unke:2021} and
references therein). Such ML-based energy functions were also extended
to reactive systems in the gas phase,\cite{MM.ht:2020} for simulations
in solution\cite{MM.fad.sol:2022} and on
surfaces,\cite{jiang.hcl:2018} and recently they were also presented
for composite systems in the gas phase to study complex combustion
processes.\cite{zhang:2020}\\

\noindent
Computational investigations of surface reactions are of particular
interest from a technological perspective. A substantial amount of
heterogeneous catalytic reactions are performed in chemical industry
and make up an important economic factor.\cite{conceptscatal:2017} The
focus of theoretical and experimental investigation are not only on
formation and breaking of chemical bonds but also the diffusion
processes as well as the prediction of scattering
experiments.\cite{kleyn:2003, zeng:2013,miranda:2013} The systems
considered range from monocrystalline surfaces of metals, metal
oxides, minerals or ionic compounds to graphene sheets, amorphous
porous carbon\cite{zobril:2021} and water
surfaces.\cite{MM.water:2020, MM.co2.wat:2021} They can be modified by
adsorbing ultra-thin layers on metal support, using metal alloy or
nanostructuring by metal cluster or single atoms or vacancies.
\cite{haruta:1989,bernhardt:2007, zhang:2013} Of particular relevance
are surface sites with coordination-unsaturated atoms on surface
edges, kinks or vacancies which have been found to be sites of
increased reactivity.\cite{frankland1925,beck:2020} However, modeling
such surfaces with impurities from high-level electronic quantum
methods is still a challenging task, in particular if dynamics
information is sought.\\

\noindent
A combination of methods with the accuracy of a quantum mechanical
treatment with a computational performance comparable to an empirical
force field would open up possibilities to investigate reactive
processes and spectroscopic properties at a quantitative
level. Advances in this direction were acknowledged by the 2013 Nobel
Prize in Chemistry.\cite{karplus:2014,warshel:2014} However,
application of accurate, physics-based force fields for
condensed-phase simulations is still not routine. The present work
highlights specific applications from our own work and from colleagues
in the field, demonstrates the opportunities of such approaches, and
discusses future prospects and reasons which slow down more rapid
adaptation of such methods in a broader sense. This overview of the
field of quantitative MD simulations focuses on applications to
questions arising in physical chemistry and biophysics. In the current
context, the quantitative aspect is judged as from comparing with
available experimental data. Therefore, the close relationship between
experimental and computational characterization of the systems is
essential.

\section{Computational Methods for Quantitative MD}
Following the time evolution of a chemical system by means of
atomistic simulations is in principle possible through {\it ab initio}
MD (AIMD) simulations. However, computational feasibility and the
limited accuracy of density functional theory (DFT) methods for
certain applications, such as reactions, often preclude using such an
approach for quantitative studies. Production runs for full AIMD
simulations are typically limited to tens or hundreds of trajectories
with simulation times up to hundreds of picoseconds at the
semiempirical or DFT level.\cite{MM.amm:2002,gerber:2006,gerber:2018}
On the other hand, empirical force fields provide a computationally
efficient way to determine the total energy and forces of systems in
the condensed phase to carry out MD simulations. However, such
(parametrized) representations are often not sufficiently accurate for
quantitative comparisons or even predictions.\\

\subsection{Potential Energy Surfaces for Gas- and Condensed-Phase Processes}
With the advent of computationally efficient and accurate electronic
structure calculations, empirical potential energy surfaces (PESs) for
small molecules can now be replaced with more accurate
representations. It is now possible to determine energies for $\sim
\mathcal{O}(10^{4})$ geometries at the coupled cluster singles/doubles
and perturbative triples (CCSD(T)) or at the multi reference
configuration interaction (MRCI) levels of theory. This shifted the
main problem to representing this information such that the PES can be
evaluated efficiently and with comparable accuracy as the underlying
quantum chemical calculations. With this, even ``spectroscopically
accurate'' calculations are now possible for small
molecules.\cite{tennyson:2016}\\

\noindent
For small molecules, in particular atom-diatom or diatom-diatom van
der Waals complexes, expansions in terms of products of Legendre
polynomials $P_{\lambda}(\cos{\theta})$, single $Y_{l,m}(\theta,\phi)$
or coupled $Y_{l1,l2,m}^{l}(\theta_1,\theta_2,\phi)$ spherical
harmonics together with radial functions $V(R)$ is
convenient.\cite{H90Ann,avoird:1994} Such an approach was used
together with explicit fits to experimental data or to reference
electronic structure calculations. Alternatively, for the long-range,
distance-dependent part explicit electrostatics using experimentally
determined or computed atomic and molecular multipoles and
polarizabilities and
hyperpolarizabilities\cite{avoird:2017,MM.heh2:2019} provides very
accurate PESs.\\

\noindent
Alternatively, permutationally invariant polynomials (PIPs) can be
used to describe the total PES.\cite{bowman.irpc:2009} This has the
added benefit that it lends itself to more straightforward
generalization to globally reactive PESs as has been done to study
reactive collisions for N$_2$ + N$_2$ $\rightarrow$ N$_2$ + 2N and
N$_2$ + N$_2$ $\rightarrow$ 4N.\cite{candler:2014} PIPs use a basis of
Morse-type functions and fit products of such basis functions to the
reference electronic structure data. Recently, PIPs have been used for
systems as large as N-methyl acetamide\cite{bowman.nma:2019} or
tropolone.\cite{bowman:2020}\\

\noindent
PESs can also be represented by kernel-based approaches such as a
reproducing kernel Hilbert space
(RKHS).\cite{rabitz:1996,hollebeek.annrevphychem.1999.rkhs,MM.rkhs:2017}
By construction, the RKHS reproduces the reference energies from
electronic structure calculations exactly on the grid points. In
addition, the physical long-range decay for large separations
$\rightarrow \infty$ can be explicitly included in the functional
dependence of the kernel. Such RKHS-based representations have been
used for entire PESs\cite{MM.cno:2018} or in a QM/MM-type approach to
treat part of an extended system with higher
accuracy.\cite{MM.cco:2013,MM.mbno:2016,MM.trhbn:2018}\\

\noindent
Finally, recent efforts for high-accuracy representations of energy
functions for individual molecules have revolved around machine
learning techniques including neural networks
(NNs),\cite{schnet:2018,MM.physnet:2019} PIPs combined with
NNs,\cite{jiang2016potential} Gaussian processes,\cite{guan:2018} or
kernel-based methods.\cite{fchl:2020,MM.h2co:2020} Recent reviews
provide a concise status of this
field.\cite{unke:2021,manzhos:2020,MM.cr:2021,behler:2021,jiang:2020,krems.2:2019,bowman:2018}\\

\noindent
For condensed phase simulations accurate electrostatic and van der
Waals interactions are essential. The most widely adopted
approximation for electrostatics is to represent the molecular charge
distribution as a superposition of point charges located on the
atoms.\cite{mackerell2004} However, such an approach neglects the
local anisotropy of the electrostatic potential (ESP) which can be
particularly relevant for halogens or charged groups. Specifically for
halogen atoms the $\sigma-$hole requires particular
attention.\cite{Clark2007sigmahole,MM.impact:2016} For such problems,
and for more generally representing a molecule's ESP in a realistic
fashion, multipolar\cite{MM.mtp:2012,MM.mtp:2013} or distributed
charge models (DCM)\cite{MM.dcm:2014,MM.dcm:2017} were developed over
the past 20 years. A non-exhaustive list of notable efforts in this
direction are the ``Sum of Interactions Between Fragments Ab Initio''
(SIBFA),\cite{piquemal:2007} ``atomic multipole optimized energetics
for biomolecular applications'' (AMOEBA),\cite{ren:2013} and the
``atomic multipole'' (MTP)\cite{MM.mtp:2016} approaches. In addition,
energy functions accounting for polarizability have been developed
which also allow to obtain more realistic, physics-based models, in
particular for condensed-phase
simulations.\cite{ren:2013,mackerell:2016}\\

\noindent
The charge distribution $\rho({\bf x})$ also depends on the geometry
${\bf x}$. Developing charge models capturing this variation can be
challenging. For one, fluctuating point charges based on the charge
equilibration scheme have been developed for this
purpose.\cite{berne:1994,mackerell:2009} They have been mainly applied
to simulate the structure and diffusivity of water, ions or amides in
water but not for infrared (IR) spectroscopy. Recently, a charge
equilibration scheme using a high dimensional NN to learn chemical
hardness and electronegativities was presented.\cite{ko:2021}
Furthermore, generalizations to simulating larger molecules are
difficult and no such extensions are available for multipolar charge
models although their conformational dependence has been
investigated.\cite{stone:1995} For CO in myoglobin\cite{Plattner08}
and CN$^{-}$ in water\cite{MM.cn:2013} conformationally dependent MTPs
have demonstrated to perform very well. For larger molecules than
diatomics fluctuating charge models have been reported by using ML
methods to predict point charges for an ensemble of structures to
match their molecular dipole moment, e.g. within the PhysNet neural
network approach.\cite{MM.physnet:2019}\\

\subsection{Dynamics and Reactions on Surfaces}
The two established ways for computations involving surfaces are by
periodic and cluster embedding models. The first model uses the
periodicity of a unit cell to describe the electronic wave function of
an infinite system. Periodic models allow meaningful computations for
metal surfaces. Care has to be taken to minimize finite size effects
such as self interaction of surface modifications or reactive sites.
Increasing the size of the unit cell lowers the self interactions but
leads to higher computational cost.\cite{illas:2005} As realistic
systems contain at least $\mathcal{O}(10^2)$ atoms, simulations for
reactions typically use generalized gradient approximation (GGA)
functionals.\cite{kroes:2021} For non-reactive systems also meta-GGA
and hybrid density functionals were already
applied.\cite{kroes.hcl:2021, ernzerhof:2003} However, the accuracy
can be increased by the specific reaction parameter (SRP) approach to
density functional theory that optimizes the mix of exchange and
correlation energies.\cite{truhlar:1999} Surface systems are also
modeled by a limited cluster surrounded by a sufficiently sized grid
of point charges corresponding to the charge of cluster atoms or
ions. Cluster embedding is usually restricted to insulators and semi
conductors but is well suited for modelling surface impurities,
supported surface clusters or reactants of low
concentration.\cite{sherwood:2001, sauer:2009, toepfer:2021} Thus,
with the cluster embedding method even coupled cluster calculations
can be carried out.\cite{sauer:2009}\\

\noindent
As for MD simulations in general, the quality of the potential energy
surface but also the validity of the Born-Oppenheimer approximation
for the surface process significantly determines the accuracy of the
simulation.\cite{saalfrank:2013} The accuracy of potential energy
surface is important for adiabatic processes such as elastic
intramolecular energy redistribution within the adsorbate when
interacting with the surface and inelastic energy transfer between
adsorbate and surface.  Experimentally, adsorbate-surface interaction
can be probed by state-to-state scattering
experiments\cite{saalfrank:2010, schafer:2015} or time dependent
desorption.\cite{ertl:2003, luntz:2005, ostrom:2013}\\

\noindent
Non-adiabatic effects on the dynamics must be considered for
simulations especially for processes on metal surfaces. Here, one
``weakly'' non-adiabatic effect is called ``electronic friction'' that
is the coupling between moving adsorbate atoms with a manifold of
electronic states of the surface via electron-hole pair transitions.
It is labelled ``weakly'' as the effect can be treated perturbatively
as a damping force on the adsorbate
movement.\cite{tully:1995,saalfrank:2013,subotnik:2018:perspective,subotnik:2018:pccp}
Additionally a random force applies on the adsorbate for surfaces of
finite temperatures.  Effects on the dynamics by coupling between two
or more PES of multiple electronic states are accounted for in the
``strong'' non-adiabatic regime.\\

\section{Vibrational Spectroscopy and Dynamics in Solution}
Vibrational spectroscopy is particularly suitable for quantitative
comparisons between experiments and simulations.  This is primarily
due to the precision with which such (laser-based) experiments can be
carried out. In the context of experiments, atomistic simulations also
yield positions and velocities of all atoms at all times. This
provides necessary information to relate experimental observables,
such as the IR spectrum, with specific structural features of a
system. Even small peptides at ambient conditions can sample multiple
conformations each of which exhibits potentially conformer-specific IR
spectra. This applies to both, peptides in the gas phase and in
solution.\cite{zwier:2006,rizzo:2009,amadei:2010} Thus it is of
interest to determine a) what conformational substate leads to a
particular spectroscopic response and b) whether there exists a unique
correspondence (``fingerprint'') between structure and
spectroscopy. One possibility is to use extensive electronic structure
calculations. This is, however, time consuming and usually only
possible for gas-phase systems.\cite{rizzo:2009,zwier:2012}
Alternatively, MD simulations with physics-based and improved force
fields can be used to determine the underlying structural features by
comparing computed and experimentally measured IR
spectra.\cite{Tokmakoff2018,MM.ala3:2021}\\

\noindent
N-methyl acetamide (NMA) is a topical example for which substantial
experimental and computational work has been carried out. In a recent
effort the frequency correlation function for NMA in water was
determined from an energy function based on reproducing kernel Hilbert
space representation for the [CONH] moiety and atomic multipoles up to
quadrupole for the electrostatics.\cite{MM.jcp:2020} Depending on the
technique used to determine the amide-I stretch frequency (scanning
along the -CO local mode or along the CONH normal mode) the three time
scales $\tau_1$ to $\tau_3$ for the decay of the frequency-frequency
correlation function (FFCF) were [0.02; 0.21; 1.00] ps or [0.02; 0.20;
  0.81] ps compared with two time scales from experiments [(0.05 to
  0.1); 1.6] ps and [0.01; 1.0] ps\cite{Woutersen:2002,Decamp:2005}
and [0.06;0.66] ps from simulations\cite{Decamp:2005} from using a
standard force field. The simulation results using a multipolar
representation favour a larger value for the decay times which is more
consistent with the experimental findings.\\

\noindent
A slightly larger and equally well-studied system is trialanine
(Ala$_3$).\cite{woutersen:2000,Woutersen,Hamm.pnas.2001,Schweitzer-Stenner.jacs.2001,Woutersen:2001,Mu.jpcb.2002,Graf2007,Gorbunov2007,Oh2010,Xiao2014,Tokmakoff2018,MM.ala3:2021}
Most experiments were carried out under conditions that prefer the
cationic species and agree that the conformational ensemble is
dominated by the poly-proline II (ppII) structure, often together with
some population of the $\beta-$sheet conformation and rare sampling of
a right-handed $\alpha-$helical structure. From 1-d and 2-d IR
experiments in the amide-I region a notable study used spectroscopic
data to refine the underlying conformational ensemble generated from
MD simulations. For this, the experimentally measured IR spectra was
reproduced best from Bayesian ensemble refinement\cite{Tokmakoff2018}
which effectively reweights a reference distribution with
corresponding conformer-specific IR spectra. Interestingly, the final
ensemble generated from such an approach was similar to the
Ramachandran maps from explicit MD simulations using a multipolar
representation which also correctly described the IR
spectroscopy.\cite{MM.ala3:2021}\\

\noindent
IR spectroscopy can also be used to follow protein assembly and
disassembly. Insulin binds to the insulin receptor in its monomeric
form but is stored in the body as a zinc-bound hexamer each of which
consists of three homodimers. Hence, the stability of the dimer to
decay into two monomers is a physiologically relevant property. For
human WT insulin the experimentally determined stabilization of the
dimer with respect to two separated monomers is $\Delta G = -7.2$
kcal/mol.\cite{strazza:1985} Thermodynamic stabilities from free
energy simulations are between $-8.4$ and $-11.9$ kcal/mol and
simulations along the minimum energy path yield $-12.4$
kcal/mol.\cite{MM.insulin:2005,MM.insulin:2018,bagchi:2019} For
pharmaceutical applications modified insulins have been synthesized
but their dimerization free energies are unknown and challenging to be
determined by standard experimental protocols. Hence, alternative
means to assess the thermodynamic stability of insulin dimer are
required. One possibility is to use infrared spectroscopy because the
insulin dimerization interface involves breaking of several hydrogen
bonds involving contacts with the -CO unit that give rise to amide-I
spectra.\cite{ganim:2010,tokmakoff:2016} Atomistic simulations using
multipolar force fields confirm that changes in the association state
result in modified amide-I spectra.\cite{MM.insulin:2020} Based on
this, attempts can be made to relate changes in the spectroscopic
response with the thermodynamic stability of insulin dimer.\\

\begin{figure}
 \begin{center}
 \resizebox{0.99\columnwidth}{!}
           {\includegraphics[scale=0.1,clip,angle=0]{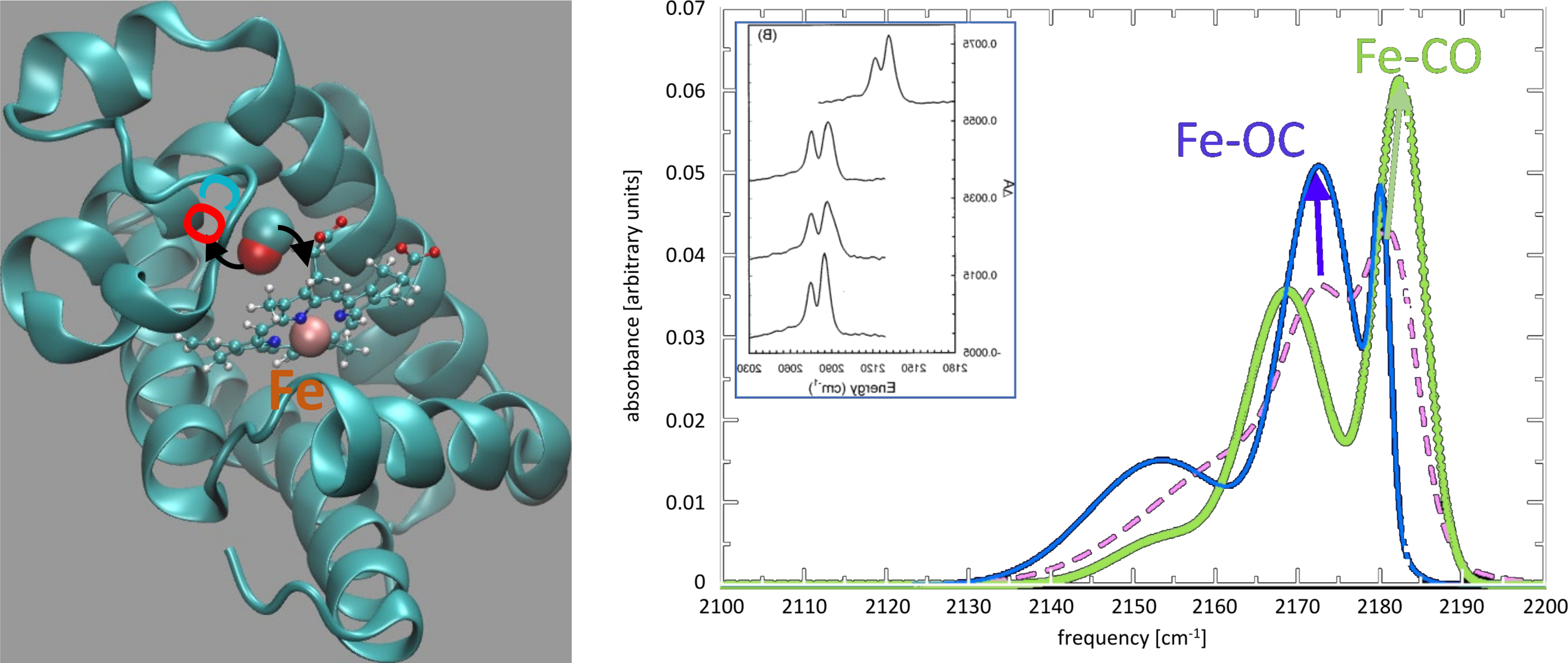}}
           \caption{Photodissociated CO and its infrared spectrum in
             Myoglobin (Mb). Left panel: Mb secondary structure in
             green with heme (ball-and-stick) and CO (van der Waals
             spheres) and the heme-Fe (green). The black arrows
             indicate rotation of the photodissociated ligand in the
             active site. Right panel: The IR spectrum for
             photodissociated CO from simulations with a multipolar
             representation of the electrostatics. The spectrum from
             the entire trajectory (dashed line) is compared with the
             spectrum from parts of the trajectory sampling the Fe--CO
             (green) and the Fe--OC (blue) substates. The inset shows
             experimental spectra for $^{12}$CO in human Mb
             (glycerol/water), $^{13}$CO in human Mb (D$_2$O),
             $^{13}$CO in sperm whale Mb (D$_2$O), and $^{13}$CO in Hb
             (D$_2$O).\cite{Lim95b} Reproduced with permission from
             Ref.\cite{Lim95b}}
\label{fig:fig2}
\end{center}
\end{figure}

\noindent
The IR spectroscopy and reaction dynamics of diatomic ligands -
including CO and NO - bound to and within Myoglobin (Mb) have been
thoroughly investigated.\cite{frauenfelder:2003} Most
characteristically, the CO infrared spectrum is split with the two
peaks separated by $\sim 10$ cm$^{-1}$. The spectroscopic signatures
were associated with two distinct conformational substates but their
structural assignment remained elusive despite dedicated
efforts.\cite{meller:1998,anselmi:2007} MD simulations with
sufficiently detailed electrostatic models for photodissociated CO
provided the necessary accuracy\cite{MM.mbco:2003,MM.mbco:2008} and
together with experimental mutation studies\cite{nienhaus:2005} the
more red-shifted peak was associated with the Fe--OC orientation
whereas the less red-shifted peak corresponds to the Fe--CO
state.\cite{MM.mbco:2006} Figure \ref{fig:fig2} shows the structure of
unligated CO in Mb together with the IR spectrum of the free CO ligand
(dashed line).  State-specific spectra for the two conformational
substates (Fe--CO in green; Fe--OC in blue) are shown as solid
lines. It is found that the total spectrum, which is the experimental
observable, agrees favourably with the measured spectrum and that the
substate-specific spectra allow assignment to an Fe--CO and Fe--OC
motif. However, although the spectra provide an identification of the
two states, the results also imply that while sampling the Fe--CO
conformation the spectroscopy is still sensitive to the presence of
the Fe--OC state and vice versa. This is due to the low isomerization
barrier between the two states. They are separated by a barrier of
$\sim 0.7$ kcal/mol which is close to the experimentally reported
barrier of 0.5 kcal/mol.\cite{kriegl:2003}\\

\noindent
Quantitative simulations are also possible for thermodynamic
properties, such as pure liquid densities or hydration free
energies. Such validated parametrizations are required for
protein-ligand binding studies, for a molecular-level understanding of
chromatography, or for reactions in solvents other than water. Recent
progress has been made in the automated parametrization for a library
of 430 molecules by focusing on the electrostatic, Lennard-Jones,
torsional and 1-4 nonbonded interactions.\cite{roux:2018} For the
density of the pure liquid the experimental reference values were
reproduced with an averaged unsigned error of 1.8\% whereas for the
enthalpy of vaporization it is 5.9\% which both are considerable
improvements over earlier
parametrizations.\cite{case:2004,mobley:2007} Nevertheless, to remove
systematic differences ($\sim 2$ kcal/mol) between experimental and
computed hydration free energies (which were determined with the TIP3P
model for water) required introducing an overall solute-solvent
scaling factor to increase the solute-solvent by 15\% on
average.\cite{roux:2018} Following a similar optimization protocol for
the polarizable Drude model the errors on the pure liquid density and
heats of vaporization were 2 \% and 6 \%, respectively, whereas for
the hydration free energies the error decreased to 0.5 kcal/mol
together with the SWM4 water model.\cite{roux:2021,swm4:2006}\\

\noindent
One recurring theme in assessing the quality of (empirical) energy
functions is that a particular parametrization may yield favourable
comparison for only one or a few experimentally determined properties
(e.g. pure liquid density and vaporization energy) - even for a class
or library of compounds - but not yield satisfactory results for other
observables (hydration free energy).\cite{roux:2021} Such
transferability has been explicitly considered for cyanide in
water.\cite{MM.cn:2011,MM.cn:2013,MM.cn2:2013} Using one single
parametrization based on fluctuating multipolar
electrostatics\cite{MM.mtp:2013} it was possible to obtain
quantitative results for vibrational energy relaxation, the 1d- and
2d-IR spectroscopy, and the hydration free energy. The computed $T_1$
times were $T_1 = 22 \pm 2$ ps and $68 \pm 11$ ps in H$_2$O and
D$_2$O, respectively, compared with $28 \pm 7$ ps and $71 \pm 3$ps
from experiments with $T_1^{\rm H_2O} / T_1^{\rm D_2O} = 0.33$ vs.
$T_1^{\rm H_2O} / T_1^{\rm D_2O} = 0.39$.\cite{MM.cn:2011,hamm:1997}
For the 1d- and 2d-infrared spectroscopy the computed blue shift was
35 cm$^{-1}$ compared with 44 cm$^{-1}$ from experiments, the full
width at half maximum of 13 cm$^{-1}$ vs.  15 cm$^{-1}$ from
experiment, and the tilt angle depending on waiting time agreed
between experiment and simulations.\cite{MM.cn:2013,hamm:1997} The
computed hydration free energy of $-76$ kcal/mol compared with $-72$
to $-77$ kcal/mol from experiment.\cite{MM.cn2:2013,pearson:1986}
Additional improvements can be obtained from using PES-morphing
techniques.\cite{MM99:nehf,bowman:1991}\\

\noindent
For pure water a range of force fields has been developed in the
recent past. One of them is the AMOEBA model which uses atomic
multipoles.\cite{amoeba:2003,iamoeba:2013} The iAMOEBA parametrization
is very successful for a wide range of $\sim 30$
properties\cite{iamoeba:2013} for which the majority of the computed
values are within a few percent of the experimentally reported
data. The E3B model follows a different strategy by adding explicit
three-body terms, akin to a many body expansion.\cite{e3b:2008}
Similarly, the HBB (Huang, Braams, Bowman) force field also uses a
many-body expansion.\cite{hbb:2006} Finally, the most comprehensive
water force field in various phases is probably the MB-Pol model which
also builds on multipolar interactions and many-body
polarization.\cite{mbpol:2013}\\

\section{Reaction Dynamics}
\subsection{Reactions in the Gas Phase}
One recent field which has witnessed quantitative simulations concerns
small-molecule reactions relevant to hypersonics and atmospheric
re-entry.\cite{sarma:2000,cummings:2003,dsmc:2017,MM.hyperson:2020}
For this, a simulation strategy involving high-level electronic
structure calculations, global and reactive RKHS-based PESs and QCT
simulations has been successfully developed and used for a range of
atom+diatom reactions.\cite{MM.cno:2018,MM.no2:2020,MM.co2:2021} The
focus was primarily on computing thermal reaction rates, final state
distributions and vibrational relaxation times over a wide temperature
range, up to 20000 K. A typical PES for one electronic state is based
on $~10^4$ energies and the reference energies are typically
reproduced within a few cm$^{-1}$ by the RKHS interpolation. Computed
thermal and vibrational relaxation rates agree to within a few percent
with those measured experimentally which is a good basis for
developing more coarse grained
models.\cite{MM.nnsig:2019,MM.nn:2020,MM.nn:2021} \\

\begin{figure}
 \begin{center}
 \resizebox{0.99\columnwidth}{!}
           {\includegraphics[scale=0.1,clip,angle=0]{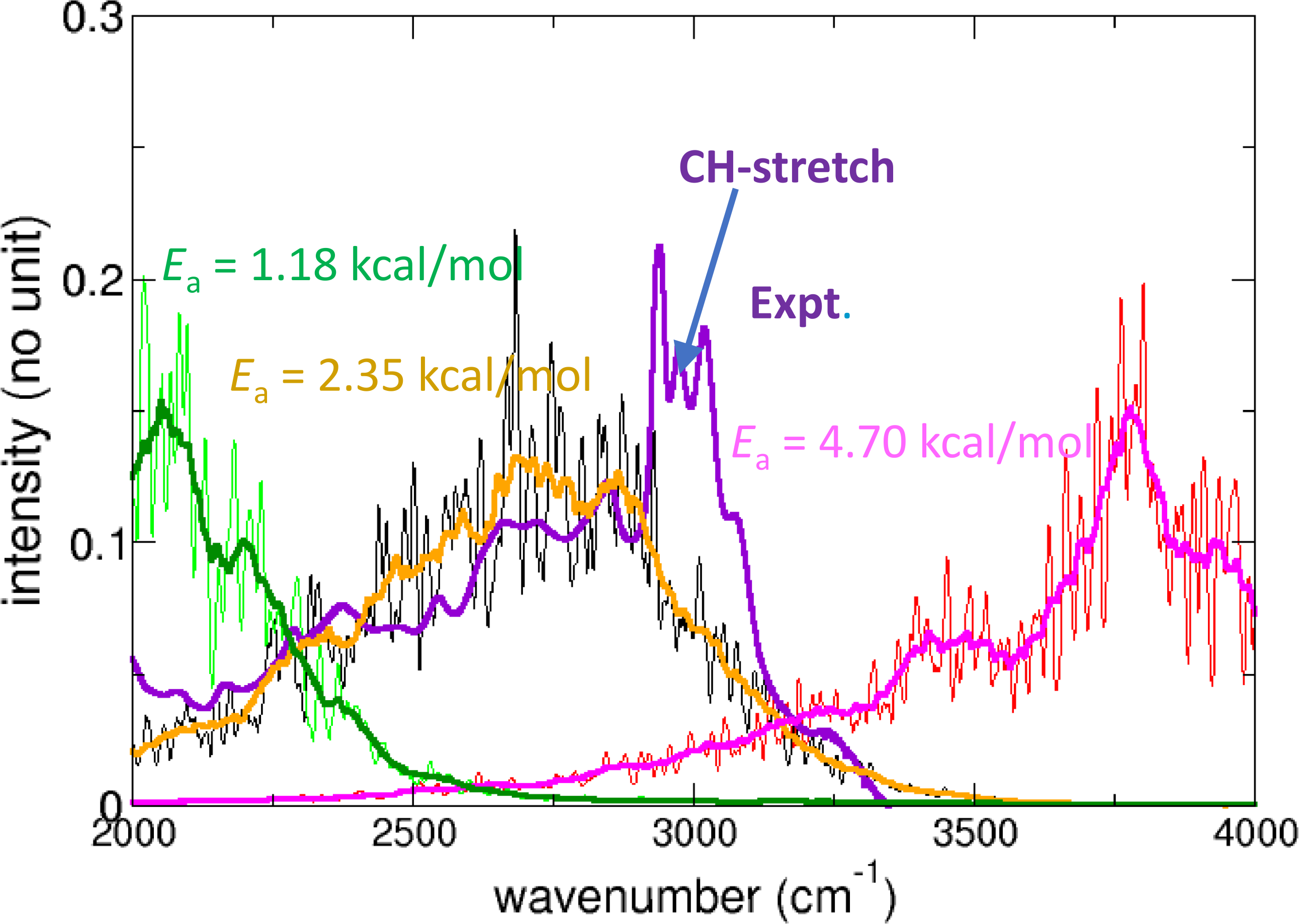}}
           \caption{Experimental infrared (blue) and computed (green,
             black/orange, red) power spectrum in the region of the
             H-transfer mode of acetylacetone. For the computations
             PESs featuring different barrier heights (1.18, 2.35,
             4.70) kcal/mol were used in a morphing-type
             approach\cite{MM99:nehf} to assess the position of the
             proton transfer band. Best agreement between
             experimentally measured and computed IR spectra is for a
             barrier height of 2.35 kcal/mol, compared with 3.2
             kcal/mol from CCSD(T)
             calculations.\cite{howardmeuwly.jpca.2015.mmpt} The
             signatures around 3000 cm$^{-1}$ in the experimentally
             measured spectra are due to the CH stretch vibrations.}
\label{fig:acac}
\end{center}
\end{figure}

\noindent
Malonaldehyde (MA), acetylacetone, and formic acid dimer (FAD) are
topical systems for quantitative simulations of gas-phase spectroscopy
and reactions. In particular MA and FAD have attracted considerable
interest and quantitatively accurate results are available for
tunneling splittings. The experimentally determined\cite{leopold:1991}
splitting for MA is 21.6 cm$^{-1}$ which compares with 23.8 cm$^{-1}$
from MCTDH calculations and 21.6 cm$^{-1}$ from Monte Carlo
simulations on the same full dimensional PES.\cite{tew:2008} For
acetylacetone, which is related to MA by substituting hydrogen atoms
by methyl-groups, the IR spectroscopy is of particular interest as the
barrier for proton transfer is low and leads to signatures in the
spectra.\cite{howardmeuwly.jpca.2015.mmpt}
Morphing\cite{MM99:nehf,bowman:1991} a parametrized PES suitable for
following proton transfer and comparing the resulting infrared
spectrum with that from experiments yields an estimated barrier for
proton transfer of 2.35 kcal/mol; see Figure
\ref{fig:acac}. Subsequent machine learning found a barrier of 3.25
kcal/mol from transfer learning to the PNO-LCCSDT(T)-F12 level of
theory.\cite{MM.ht:2020}\\

\noindent
For FAD the gas phase infrared spectroscopy contained signatures for
double proton transfer (DPT).\cite{mackeprangmeuwly.pccp.2016.mmpt}
Comparing computed IR spectra from MD simulations on a morphed
PES\cite{MM99:nehf} with those measured experimentally yielded an
estimated barrier for DPT of 7.2
kcal/mol\cite{mackeprangmeuwly.pccp.2016.mmpt} which compares with 7.3
kcal/mol from a subsequent analysis of microwave
spectra\cite{caminati:2019} and 8.2 kcal/mol from fitting CCSD(T)-F12a
energies calculated with the cc-pVTZ and aug-cc-pVTZ basis sets for H
and C/O atoms (CCSD(T)-F12a/haTZ) to permutationally invariant
polynomials.\cite{bowman.fad:2016} A recent full dimensional PES which
was transfer learned from the MP2 level of theory to CCSD(T) energies
reported\cite{MM.fad:2022} a barrier for DPT of 7.92 kcal/mol and a
dissociation energy for FAD in the gas phase from diffusion Monte
Carlo simulations of $D_0 = -14.23 \pm 0.08$ kcal/mol in excellent
agreement with an experimentally determined value of $-14.22 \pm 0.12$
kcal/mol.\cite{suhm:2012} These examples illustrate the benefit of
accurate, reactive and full-dimensional PESs for quantitative
simulations of gas phase processes.\\

\noindent
QCT simulations were also applied to the photodissociation of
formaldehyde (H$_2$CO) following excitation of the 2$^1$4$^3$
band. The results from QCT simulations on a PIP-represented PES
determined at the MRCI/cc-pVTZ level of theory compared with
experiments yielded excellent results.\cite{houston:2017} The
properties considered included the speed distributions for
state-selected CO, and the rotational and vibrational distributions of
the CO and H$_2$ products.\cite{houston:2017} In a more recent study
the kinetic energy release of the three-body breakup from Coulomb
explosion experiments was compared with that from QCT simulations. For
high kinetic energy particularly good agreement was found whereas for
lower energies the agreement was rather more
qualitative.\cite{endo:2020}\\

\noindent
Atomistic simulations can also be successfully used for determining
vibrational relaxation rates. This was done\cite{chen:2020} for the
CO+CO collision based on an accurate 6-dimensional PES (frozen
monomers) validated for 68 rovibrational levels with a root mean
square error of 0.29 cm$^{-1}$ between experiment and
calculations.\cite{surin:2007} The vibrational relaxation rates from
QCT simulations for the $v=10$ and $v=16$ vibrational levels were $2.6
\times 10^{10}$ cm$^3$/s and $3.1 \times 10^{10}$ cm$^3$/s compared
with $1.9 \times 10^{10}$ cm$^3$/s and $5.0 \times 10^{10}$ cm$^3$/s
from experiments.\cite{deleon:1986} Such information is paramount for
combustion processes and hypersonics and was also determined in
near-quantitative agreement for the N+NO and O+CO
systems.\cite{MM.n2o:2020,MM.co2:2021} Also, for the H+O$_3$
$\rightarrow$ OH+O$_2$ reaction, QCT simulations based on a PIP+NN
representation of energies calculated at the MRCI level of theory
reported final state vibrational distributions for the OH product
peaking at $v' =8$, in near-quantitative agreement with available
experiments.\cite{chen:2021}\\

\noindent
Finally, thermal rates for reactions have also been determined for
bimolecular processes. As an example, the kinetics of the OH+HO$_2$
reaction to form O$_2$+H$_2$O was determined from QCT simulations
using a full dimensional PES based on PIP + NN.\cite{liu:2019} Over a
temperature range between 250 K and 3000 K the computed rate is in
good agreement with a range of experiments and reproduces in
particular the negative temperature dependence reported at low
temperatures.\\

\subsection{Reactions in Solution}
For reactions in solution the additional complication is the presence
of an environment that needs to be described by a separate energy
function. The most common solvent, water, is particularly challenging
to represent and despite immense effort to date no single
parametrization outperforming all others is
available.\cite{headgordon:2018} In addition, many empirical energy
functions which are a common starting point for investigating
reactions in solution have been parametrized with one of the simpler
water force fields such as TIP3P\cite{Jorgensen.tip3p:1983} or
SPC/E.\cite{spce:1987}\\

\noindent
Despite the challenges, meaningful quantitative simulations for
certain reaction types in solution are available. One of them concerns
the Claisen rearrangement reaction.\cite{claisen:1912} Using multi
state adiabatic reactive dynamics (MS-ARMD) simulations, the
transformation from chorismate to prephenate and corresponding smaller
molecular systems was investigated.\cite{MM.claisen:2019} The
catalytic effect of the protein over the reaction in water was
correctly captured whereas the actual activation free energy did not
reproduce that experimentally observed. This can be achieved from
using a dedicated parametrization as was done for EVB-based
simulations of the same reaction.\cite{warshel:2003} With a model
parametrized to reproduce the activation free energy in water the
barrier height in the protein environment was reproduced to within 0.6
kcal/mol compared with experiment.\\

\noindent
Another class of reactions that has been investigated in great detail
in solution are S$_{\rm N}$2 reactions. One example is the S$_{\rm
  N}$2 reaction of haloalkan dehalogenase (DhlA) in which the
nucleophile carboxylate from Asp124 replaces one of the halides of the
CH$_2$Cl-CH$_2$Cl substrate, i.e. --COO$^-$+(CH$_2$Cl)$_2$
$\rightarrow$ --OCO-CH$_2$-CH$_2$Cl + Cl$^-$.\cite{warshel:2004}
Following the empirical valence bond (EVB)
approach,\cite{warshel:1980} first the reference reaction in water was
parametrized to reproduce the experimentally observed activation free
energy. This model was then used for simulations in the protein DhlA
and the computed catalytic effect of 11.6 kcal/mol was in good
agreement with the experimentally observed value of 11.7
kcal/mol.\cite{warshel:2004} For the [Br-CH$_3$-Cl]$^-$ reaction in
solution parametrizations within the MS-VALBOND\cite{MM.msvb:2018} and
MS-ARMD\cite{nagy.jctc.2014.msarmd} frameworks have been successfully
used.\cite{MM.sn2:2019} The catalytic effect in going from the gas
phase to solution for the forward [Br-CH$_3$ + Cl]$^-$ $\rightarrow$
[Br + CH$_3$-Cl]$^-$ reaction is 17.4 kcal/mol using MS-ARMD at the
MP2 level, compared with 15.2 kcal/mol from experiment.\\

\noindent
Phosphate hydrolysis reactions have also been investigated at a
quantitative level. For a number of substituted methyl phenyl
phosphate diesters activation free energies were determined from EVB
calculations and compared with experiment.\cite{warshel:2008} Typical
differences between computed and measured activation free energies
were below 1 kcal/mol. From analysis of the MD trajectories it was
concluded that phosphate transfer in these systems followed an
associative rather than a dissociative pathway. For methyl transfer
reactions, recent MS-ARMD simulations also found that the pathway is
associative. For one of the systems, Pyr+MeBr experimentally
determined barrier heights for acetonitrile and hexane as the solvent
were 22.5 kcal/mol and 27.6 kcal/mol which compare with 23.2 kcal/mol
and 28.1 kcal/mol from MS-ARMD simulations.\cite{MM.ch3:2021}\\

\subsection{Surface Reactions}
Molecular beam scattering experiments are an important tool to
investigate the reaction dynamics at the gas-surface interface and
yield information about reaction mechanisms, topological features and
the gas-surface interaction potential. Their characterization and
accurate prediction of suitable observables such as the final
vibrational, rotational or translational energy distribution of the
reaction products, reaction rates, or desorption energies from
simulations is of great importance not only for understanding but also
for designing new heterogeneous
catalysts.\cite{kroes:2002,freund:2005,wodtke:2016a,wodtke:2016b} \\

\noindent
One prominent process is the scattering of NO from a metal Au(111)
surfaces that is focus of research for the past 20
years.\cite{wodtke:2000} The scattering of initially highly
vibrationally excited NO molecules ($\nu_{\rm ini}=15$) has shown
vibrational relaxation of about $36$\,kcal/mol ($150$\,kJ/mol) or up
to 10 vibrational quanta. This differs significantly for NO scattering
from insulators such as LiF(001) for which negligible vibrational
energy loss was observed which is indicative of non-adiabatic
relaxation channels rather than purely mechanical
ones.\cite{wodtke:2000,auerbach:2003}\\

\noindent
Early simulations by Tully and coworker based on the independent
electron surface hopping (IESH) model allow metal-to-molecule charge
transfer by including two electronic states in the model
Hamiltonian.\cite{tully:2009a,tully:2009b} Such an approach only
qualitatively captures the large experimentally observed vibrational
relaxation. As an example, simulations for the scattering of
NO($\nu_{\rm ini}=16$) with an incident energy of $E_{\rm
  ini}=0.5$\,eV lead to a most probable population of the $\nu_{\rm
  fin}=14$ final vibrational state whereas experiments report
pronounced relaxation to $\nu_{\rm fin,exp}=6$.\cite{schafer:2015}
Including electron friction in the model to perturbatively describe
vibration-electron coupling yields relaxation to $\nu_{\rm
  fin}=13$.\cite{tully:1995,salin:2008,saalfrank:2010,schafer:2015}
These differences between experiment and simulations were primarily
attributed to inaccuracies in the PES which also may explain the
observed overestimation of multibounce events in the
simulations.\cite{tully:2009a,bartels:2014}\\

\begin{figure}
 \begin{center}
 \resizebox{0.99\columnwidth}{!}
           {\includegraphics[scale=0.1,clip,angle=0]{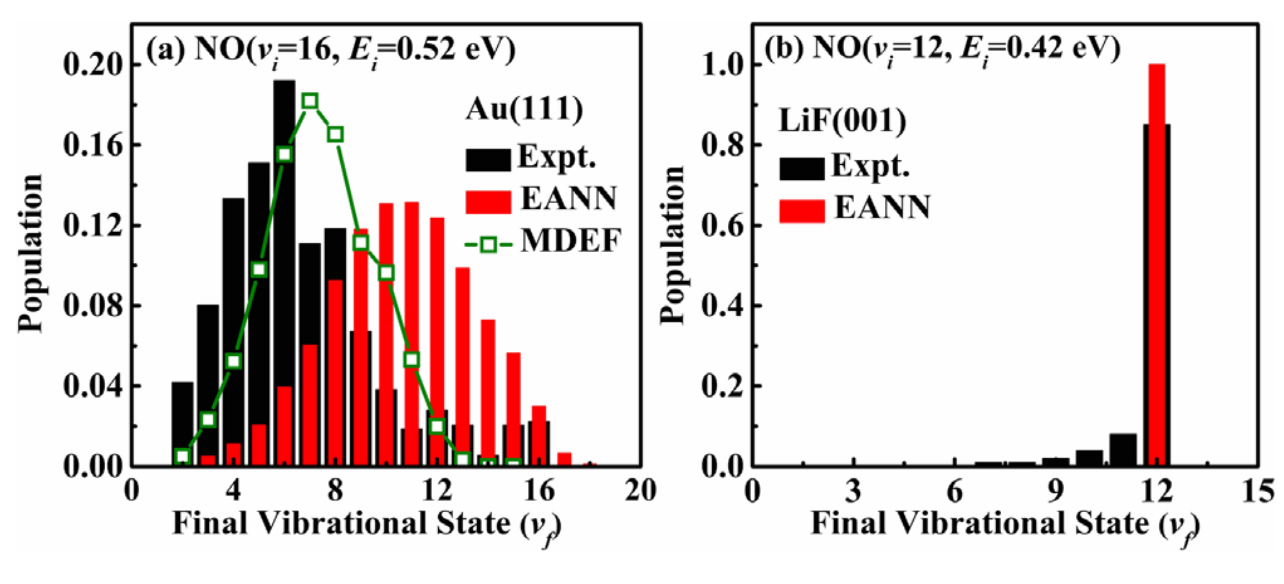}}
\caption{Final vibrational distribution of highly vibrationally
  excited NO scattered from (a) Au(111) and (b) LiF(001) from
  adiabatic (red) and non-adiabatic MD simulations including
  electronic friction (open green squares). Experimental data of are
  shown in black.\cite{wodtke:2000,auerbach:2003}. Figure from
  Ref. \citenum{jiang:2021}.}
\label{fig:fig3}
\end{center}
\end{figure}

\noindent
Figure~\ref{fig:fig3}a shows one of the most recent improvements
towards a quantitative study of NO-vibrational relaxation using
accurate energy functions in MD simulations including electron
friction (MDEF, open green squares).\cite{jiang:2021} Vibrational
relaxation of scattered NO($\nu_{\rm ini}=16$, $E_{\rm ini}=0.52$\,eV)
on Au(111) leads to a loss of $9$ quanta to yield $\nu_{\rm fin} = 7$
which is close to the experimental peak at $\nu_{\rm fin,exp} =
6$.\cite{wodtke:2000} This corresponds to a vibrational energy loss of
$\sim 23$\,kcal/mol ($\sim 1.0$\,eV) compared with the experimentally
observed $36$\,kcal/mol ($150$\,kJ/mol). Even without electronic
friction (red bars) QCT simulations with the embedded atom neural
network (EANN) potential model already predicts a considerably more
pronounced relaxation with a peak at $\nu_{\rm fin} = 11$. Conversely,
when NO($\nu_{\rm ini}=12$, $E_{\rm ini} = 0.42$\,eV) scatters from
LiF(001), the product state distribution $P(\nu_{\rm fin})$ strongly
peaks at $\nu_{\rm fin}=12$ and no vibrational relaxation takes place,
see Figure~\ref{fig:fig3}b. This finding primarily reflects the rather
isotropic interaction between NO and the LiF surface that governs the
dynamics.\cite{jiang:2021} Similarly, the predicted translational
energy loss of scattered NO($\nu_{\rm ini} = 1$, $E_{\rm ini} =
0.31$\,eV) of $\left< \Delta E_\mathrm{trans} \right> = 4.8$\,kcal/mol
($0.21$\,eV) agrees well with the experimentally determined value of
$4.4$\,kcal/mol ($0.19$\,eV).\cite{loy:1985} Hence, a quantitative
understanding of such experiments is possible.\\

\noindent
These results emphasize the need for high-dimensional and accurate
PESs to describe energy transfer between the surface and the impacting
molecules together with including non-adiabatic effects especially for
simulation on metal surfaces. In the present case the improvement is
due to using a machine-learned, reactive PES that correctly captures
the decrease in the NO dissociation barrier from $\sim 161$\,kcal/mol
($7.0$\,eV) in the gas phase to $66$\,kcal/mol
($2.86$\,eV)\cite{jiang:2021} on Au(111) which is below the
vibrational energy at $\nu = 16$ (corresponding to $78$\,kcal/mol or
$3.4$\,eV).\cite{schafer:2015,schaefer:2016,jiang:2019} The NO bond
elongation up to $1.89$\,\AA\/ leads to softening of the molecular
vibration and increased coupling with other surface degrees of
freedom.\\

\noindent
For CO on Au(111) a NN representation of the Behler-Parrinello type
fitted to DFT reference data was used together with QCT simulations to
investigate the scattering of vibrationally excited CO$(v_{\rm ini} =
2)$ from the surface. The simulation shows predominantly specular
scattering with no significant deviation between incidence angle
($9^\circ$) and scattering angle (peak at $10^\circ$), consistent with
an experimentally observed scattering angle of $9^\circ$. More
importantly, the experimentally observed CO($v_{\rm fin} = 1)$ product
could be attributed to rapid nonadiabatic relaxation of chemisorbed
CO.\cite{huang:2019} The vibrationally relaxed CO can either desorb
rapidly or transfer into a long living physisorbed state giving raise
to the fast and slow component in the experimental time of flight
spectra.\cite{shirhatti:2018}\\

\noindent
Achieving quantitative simulation of surface reactions is still
challenging but highly relevant for developing and improving
heterogeneous catalysts.\cite{guo:2019,kroes:2021,wodtke:2021}
Although the prediction of the reaction probability for the
dissociative chemisorption of molecular H$_2$ on metal surfaces is
possible with high accuracy,\cite{diaz:2016} for practical application
reliable prediction of heavier atoms are required. Neglecting surface
atom motion, simulations of H$_2$ on Cu(111) yield a reaction
probability curve that is shifted by less than 1\,kcal/mol along the
collision energy range up to $19.1$\,kcal/mol
($80$\,kJ/mol).\cite{kroes:2009} \\

\noindent
One limitation in such simulations concerns the level at which the
electronic structure calculations can be carried out due to the system
size and time scales required for meaningful simulations. The current
state-of-the art for modeling molecule-metal reactions is still DFT
with GGA functionals.\cite{kroes:2021} GGA density functionals have
shown good performance for predicting sticking probabilities of
dissociative adsorptions if the work function of the metal surface
exceeds 7\,eV.\cite{kroes:2020b} One successful example is the match
between \emph{ab initio} MD simulations of the dissociative adsorption
of CH$_4$ on Ni(111) surfaces at $550$\,K with molecular beam
experiments. The SRP approach on density functionals was used for an
accurate interpolation of the dissociation barrier
height.\cite{utz:2016, truhlar:1999} They achieved chemical accuracy
for low incident energies ranging from $24.1$\,kcal/mol up to
$28.7$\,kcal/mol ($101-120$\,kJ/mol).\cite{utz:2016} At higher
incident energies the differences between simulation and experiment
increased. It was argued that this is due to a deficiency in the
classical dynamics description to account for quasi-resonant energy
transfer. However, quantitative agreement was found with initially
excited CH-stretch vibrations ($\nu=1$) at incident energies up to
$38.2$\,kcal/mol ($160$\,kJ/mol). Subsequently, the development of a
15-dimensional PIP-based PES for CH$_4$ on Ni(111) fitted to nearly
200\,000 reference data points has opened up the possibility for
statistically quantitative description.\cite{jiang:2017} Chemical
accuracy for the dissociative adsorption of CHD$_3$ on Pt(111) by
\emph{ab initio} MD simulations was also
achieved.\cite{kroes:2018chd3,kroes:2016chd3,kroes:2014chd3} \\

\noindent
For metal surface systems with low work functions pure density
functionals perform less well for reaction barriers and sticking
probabilities of dissociatively adsorbed molecules.\cite{kroes:2020b}
One well-studied example system is the adsorption of O$_2$ on Al(111)
surfaces. Here, experiments have shown small initial sticking
probabilities of 1\% at low incident energies ($\sim 0.7$\,kcal/mol,
$30$\,meV) and an increase to $\sim 90$\% at high incident energies
($>13.8$\,kcal/mol, $600$\,meV).\cite{kasemo:1997} Adiabatic MD
simulations on PESs from GGA density functionals yield a barrier-less
O$_2$ dissociation on the Al(111) surface and predict a 100\% sticking
probability even for the lowest simulated incident energy of
$0.6$\,kcal/mol ($25$\,meV). Only energy surfaces from
spin-constrained density functional computations with O$_2$ in the
electronic triplet state yield qualitative agreement for the sticking
probabilities.\cite{scheffler:2008} Nearly quantitative agreement has
been achieved by using a PES determined from embedded correlated wave
function theory\cite{carter:2014} that treats a molecule-surface
embedded subsystem with more accurate electron correlation methods and
the extended surface at the DFT level. MD simulations were carried out
on a fitted 6-dimensional London-Eyring-Polanyi-Sato
PES\cite{conte:2009,ootegem:2010} using 700 reference points for O$_2$
on a 10 to 14 atoms embedded Al(111) slab in a periodic supercell on
MP2 level. The results yield a simulated probability sticking curve
that is about $1.1$\,kcal/mol ($0.05$\,eV) shifted towards higher
incident energies.\cite{jiang:2018} \\

\noindent
In another recent effort\cite{nandi:2021} the vibrationally induced
isomerization of CO on NaCl surfaces was investigated from QCT
simulations. Experiments have reported the isomerization of OC--NaCl
to CO--NaCl for highly vibrationally excited CO. For the largest
cluster considered, containing 13 CO molecules adsorbed onto the NaCl
surface, isomerization starts with $v=22$ quanta in the CO stretch,
consistent with experiment. Even after isomerization, the CO adsorbate
remains vibrationally excited which is also what experiments report.\\

\noindent
Quantitative MD simulations have also been possible for
chromatographic systems. The dynamics of single molecules at the
solid/liquid interface plays important roles in surface science
(adsorption/desorption), material science, and interfacial chemistry.
The molecular mechanisms underlying retention in RPLC depends on both,
specific and non-specific interactions between the analyte and its
environment.\cite{Dorsey1989rplc,Dorsey1994retention,Gritti2001inter,Wolcott2000inter,Krupczynska2004inter,Buszewski2012inter}
The non-polar stationary phase is often a functionalized silica
surface to which alkyl chains of various lengths are tethered, and the
typical mobile phase is a water/methanol or water/acetonitrile
mixture. Despite the superficially “simple” chemical composition of
such systems, a molecular understanding underlying the separation
process is challenging as the system is highly dynamical,
heterogeneous and
disordered.\cite{markus:2007,meuwly:2010,gupta2012,gupta2016chroma}
Experimentally determined capacity factors which can be converted into
retention times and free energies of binding are available for
toluene, benzene, chlorobenzene, and phenol.\cite{Buszewski1990305}
Using multipolar force fields that reproduce the hydration free
energies for the probe molecules,\cite{MM.sigma:2016} free energy
differences $\Delta \Delta G_{\rm PhCl \rightarrow \rm PhR}$ for
mutating -Cl in chlorobenzene to -H (benzene), -OH (phenol), and
-CH$_3$ (toluene) were determined inside the chromatographic
column. With a 20:80 MeOH/H$_2$O solvent mixture these computed free
energy changes for toluene, benzene, and phenol are [0.05; 0.57; 1.25]
kcal/mol compared with [0.06; 0.60; 1.43] kcal/mol from experiment and
for the 50:50 MeOH/H$_2$O mixture they are [0.08; 0.51; 1.01] kcal/mol
compared with [0.20; 0.61; 1.16] kcal/mol.\cite{MM.chroma:2017}\\

\section{Outlook}
In this outlook a number of possible future developments are
described. They include conceptual and practical aspects of force
fields and further refinements to better capture the underlying
physics of the molecules considered.\\

\noindent
The majority of atomistic simulations in solution are carried out with
rather empirical energy functions for the solvent. This is
particularly obvious for protein dynamics using
TIP3P\cite{Jorgensen.tip3p:1983} or SPC/E\cite{spce:1987} models for
the solvent which are convenient but not quantitative. One of the main
reasons for this is the fact that the corresponding energy functions
for the proteins have been parametrized with such water models early
on. Only replacing the underlying water model will lead to an
imbalance between the water-water and water-protein interactions which
is not desirable. Hence, only a rigorous ``bottom-up''
reparametrization of protein force fields using improved but still
computationally efficient water energy functions can address
this. Recently, an effort has been made to emulate a wide range of
rigid, fixed-charge and polarizable water parametrizations using the
minimally distributed charge model\cite{MM.dcm:2017} within the same
implementation.\cite{MM.dcm:2020} This can serve as a proxy for a
unified parametrization platform for small molecules and protein
fragments in solution based on improved energy functions for water.\\

\noindent
For chemical reactions in solution the reorganization of the electrons
makes an important contribution to the solute-solvent interaction. In
the gas phase such effects can be absorbed in the total energy if a
global representation of the total energy function is
possible. However, in solution, the charge reorganization needs to be
included explicitly as a function of geometry because the
solvent-solute interaction is described by explicit electrostatics in
one or the other way. A recent example concerns the recombination
dynamics of CO and oxygen atoms. For the process in gas phase all
necessary interactions are accounted for in one global energy
function\cite{MM.co2:2021} whereas for the process on amorphous solid
water the reorganization of the electrons between the CO+O reactant
and the CO$_2$ product needs to be included
explicitly.\cite{MM.co2.wat:2021,MM.co2:2022}\\

\noindent
Conformationally dependent electrostatics has already been explored to
some extent in the past. However, the applications were often
restricted to water\cite{berne:1994} with the exception of the
fluctuating charge implementation in CHARMM.\cite{patel:2004} The
model itself is based on charge equilibration between bonded atoms
based on their electronegativity and chemical hardness rather than
first principles electrostatic potentials from electronic structure
calculations. For additional accuracy, multipolar energy functions can
include conformational dependence of the MTPs which has, however, only
been done for diatomics.\cite{Plattner08,MM.cn:2013}\\

\begin{figure}
 \begin{center}
 \resizebox{0.99\columnwidth}{!}
           {\includegraphics[scale=0.1,clip,angle=0]{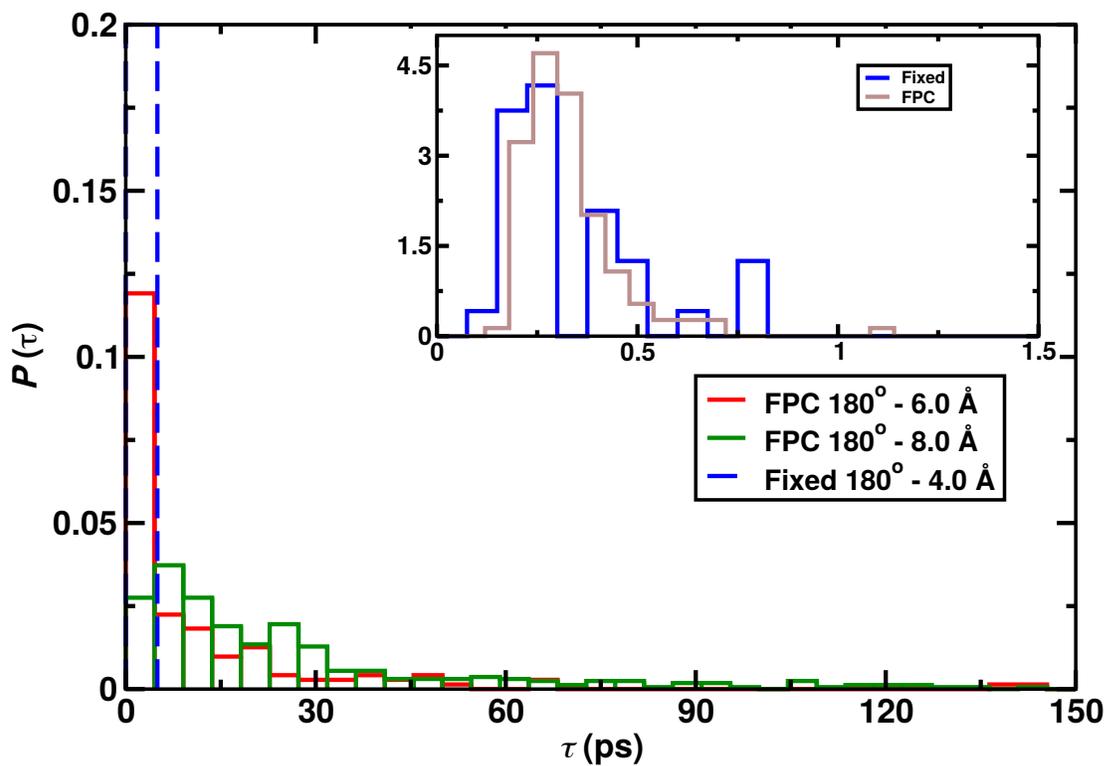}}
\caption{Normalized reaction time distributions for CO + O
  recombination from 500 simulations, for reactions initiated from
  $\theta = 180^\circ$ and $R = 4.0$, 6.0, 8.0 \AA\/ using
  fixed\cite{MM.co2.wat:2021,MM.co2:2022} and fluctuating point charge
  (FPC) models. The inset reports the distribution for $R = 4.0$ \AA\/
  and almost all recombination takes place within 1 ps. The reaction
  probability found with $R= 4.0$ \AA\/ is 32\% and 99.2\% with fixed
  and fluctuating charges respectively.}
\label{fig:co2}
\end{center}
\end{figure}

\noindent
As an example for the influence of conformationally dependent charges,
Figure \ref{fig:co2} reports the CO$_2$ formation probability for
different initial conditions of the reactive MD
simulations. Recombination simulations were initiated from a CO(center
of mass)--O separation of i) $R = 4.0$ \AA\/ for a previously
used\cite{MM.co2.wat:2021} fixed point charge model in the collinear
OC--O ($\theta = 180^\circ$) conformation and ii) for separations
ranging from $R = 4.0$ \AA\/ to $R = 8.0$ \AA\/ with a fluctuating
charge model. With fixed charges\cite{MM.co2.wat:2021,MM.co2:2022} the
oxygen atom interacts more strongly with the water hydrogen atoms due
to its charge of $q_{\rm O} = -0.2$e. Although some charge transfer
between the water surface and the oxygen atom is expected from
electronic structure calculations\cite{MM.co2.wat:2021} the magnitude
of the charge is probably too high. On the other hand, with the
fluctuating charge model it is possible to correctly describe the
change in the charges of the CO and O moieties as they approach one
another. In this model, the magnitude of the point charges depends on
the OC--O distance. Charges from reference electronic structure
calculations were found to change in a sigmoidal fashion between the
CO$_2$ product geometry and the CO+O reactant state.\\

\noindent
As is shown in the inset of Figure \ref{fig:co2}, for the shortest
separation the normalized reaction time distributions for the two
charge models are close but the total product formation probability
differs considerably. It is 32\% for fixed charges and 99\% for
fluctuating charges from running 500 simulations. As in the reactant
geometry the CO and O fragments are close to electrically neutral they
interact weakly with the surface and diffusion is facile (diffusional
barrier of 0.2 kcal/mol to 2 kcal/mol for atomic
oxygen)\cite{MM.oxy:2018} which increases the probability for reactive
encounters. Hence, recombination from short distances $R$ is almost
100\% and recombination from further away is still possible on the 100
ps time scale, see Figure \ref{fig:co2}. The desorption energy for
atomic oxygen with $q_{\rm O} = -0.2$e is $\sim 7$ kcal/mol which also
explains its slow surface diffusion with fixed
charges. Experimentally, the desorption energy ranges from 2.4 to 3.5
kcal/mol\cite{minissale:2016} compared with 2.2 kcal/mol from
simulations with neutral oxygen ($q_{\rm O} = 0$e).\cite{MM.oxy:2018}
For molecular oxygen (O$_2$) atomistic simulations
found\cite{MM.o2:2019} desorption energies of 1.5 to 2.0 kcal/mol
which compares favourably with experiments that report a value of 1.8
kcal/mol.\cite{noble:2012} In summary, with fluctuating charges
correctly describing the asymptotic states of the
OC+O$\rightarrow$CO$_2$ reaction the diffusional and desorption
barriers for the oxygen atom are in quantitative agreement with
experiment and allow to realistically model recombination on amorphous
solid water.\\

\noindent
For energy transfer and spectroscopy the coupling between internal
degrees of freedom is important. However, in most empirical energy
functions this mechanical coupling is implicit and primarily governed
by coordinate transformations between Cartesian coordinates which are
used for following the molecular dynamics and the internal coordinates
in which the energy function itself is described. There are examples
for force fields such as the Merck Molecular Force Field
(MMFF)\cite{mmff:1996}, which include cross terms, e.g. between
neighboring bonds, bonds and valence angles. However, the number of
parameters to be determined increases considerably and the additional
number of terms affects the computational performance. ML-based
techniques provide an opportunity to include such couplings from
rigorous electronic structure data and first examples demonstrate
their accuracy for chemical reactions and
spectroscopy.\cite{MM.atmos:2020,MM.fad:2022}\\

\noindent
Dynamical simulations and the chemical understanding of surface
processes are a crucial factor for innovations in heterogeneous
catalysis. The difficulties and large effort in performing accurate,
high-level quantum electronic computations together with incorporating
non-adiabatic effects during molecule-surface interactions has often
hindered quantitative agreement between simulations and experimental
observations. Increased understanding in surface effects,
sophisticated models for accurate potential energy surfaces and ML
potentials that allow the simulation of adsorbate and surface atom
dynamics have lead to increasing success of reproducing experimental
results. However, there are examples like the dissociative
chemisorption of HCl on Au(111) for which quantitative molecular
dynamics simulation could not successfully achieved
yet.\cite{guo.hcl:2016,kroes.hcl:2016,jiang.hcl:2018,kroes.hcl:2019,kroes.hcl:2020}
\\

\noindent
Quantitative atomistic simulations provide considerable scope for
molecular-level understanding of complex chemical and biological
systems. For this, computationally efficient, versatile and
implementations of sufficiently accurate representations of the total
energy of the systems are required. The ``quantitative'' nature of
such simulations is ultimately judged from comparison with experiments
which in itself has measurement errors associated with it. An
integrated approach combining knowledge from experiment and simulation
together with a realistic assessment of the uncertainties involved in
both will be of particular interest to arrive at a comprehensive
description and understanding of complex systems. In this regard,
ML-based techniques will contribute to both, the representation of the
energy function and the molecular simulations themselves, and the
quantification of the underlying uncertainties.\\
        
\section*{Acknowledgments}
This work was supported by the Swiss National Science Foundation
through grants 200021-117810, 200020-188724, the NCCR MUST, AFOSR, and
the University of Basel (to MM). This project also received funding
(to KT) from the European Union’s Horizon 2020 research and innovation
programme under the Marie Skłodowska-Curie grant agreement No 801459 -
FP-RESOMUS.

\bibliography{refs.tc}

\begin{figure}
 \begin{center}
 \resizebox{0.35\columnwidth}{!}
           {\includegraphics[scale=0.1,clip,angle=0]{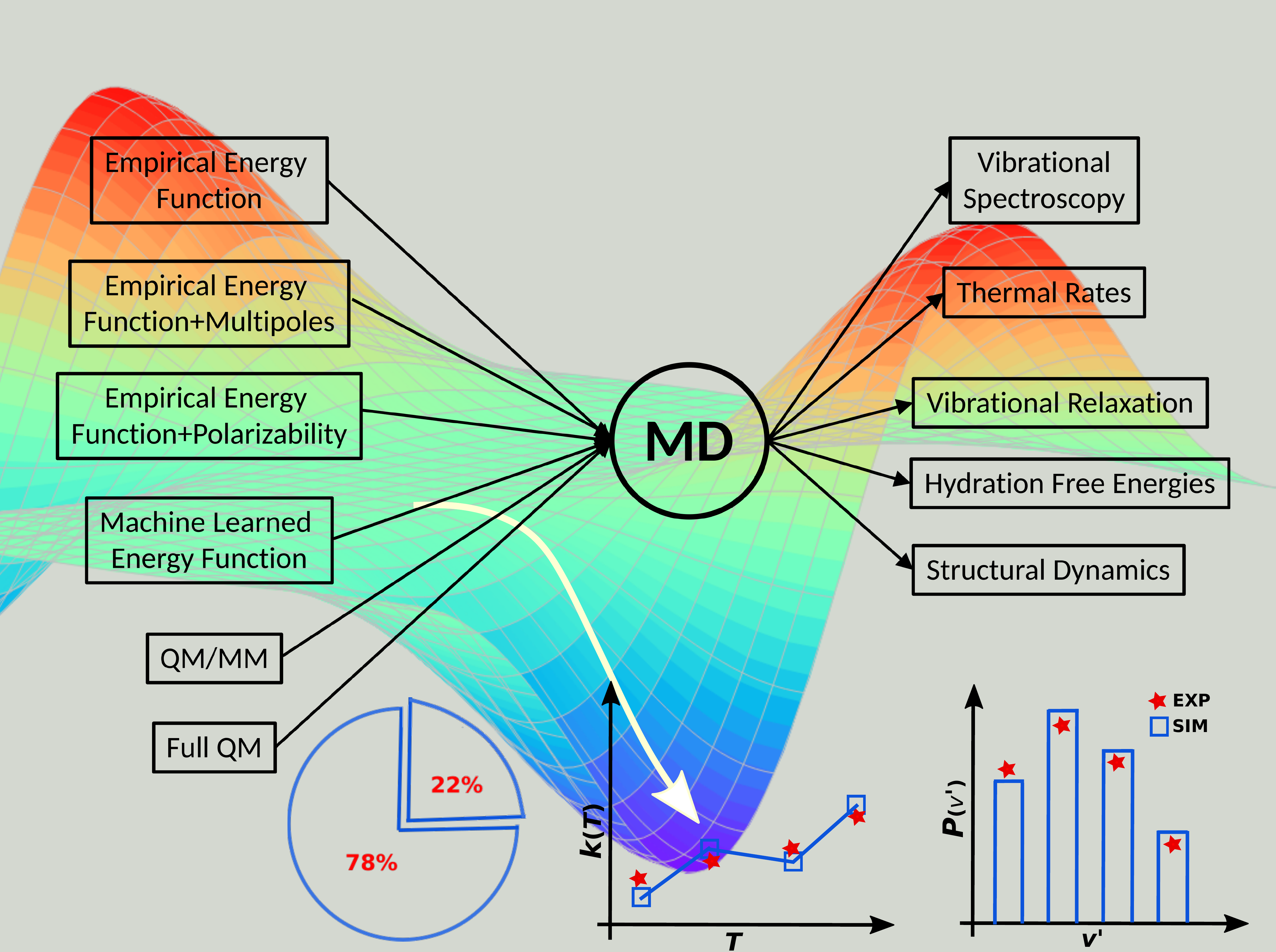}}
           \caption{Table of Contents Graphics}
           \end{center}
\end{figure}

\end{document}